\documentclass[twocolumn,showpacs,preprintnumbers,prb,amsmath,amssymb]{revtex4}
\usepackage{graphicx}
\usepackage{dcolumn}
\usepackage{bm}
\begin{document}

\title{Damage along an ion track in diamond: a computer simulation
  }

\author{ A. Sorkin }
\email{anastasy@techunix.technion.ac.il }
\homepage{ http://phycomp.technion.ac.il/~anastasy/}
\author{ Joan Adler}
\author{ R. Kalish}
\affiliation{ Physics Department, Technion, Israel Institute of Technology, 
Haifa, Israel, 32000  }

\date{\today}

\begin{abstract}

 We present tight-binding molecular dynamics
simulations of the structural modifications that result from the 
"thermal spike" that occurs during the passage of a heavy fast
ion through a thin diamond or amorphous carbon layer, and
the subsequent regrowth upon cooling. The thermal spike and cooling down
are simulated by locally heating and then quenching a
small region of carbon, surrounded either by diamond or by a mostly
$sp^3$ bonded amorphous carbon network. For the case of the thermal
spike in diamond we find that if the ``temperature'' (kinetic
energy of the atoms) at the center of the thermal spike is high
enough, an amorphous carbon region containing a large fraction of
threefold coordinated C atoms ($sp^2$ bonded) remains within the
diamond network after cooling. The structure of this amorphous
layer depends very strongly on the ``temperature'' of heating 
and
on the dimensions of the thermal spike. Scaling is found between
curves of the dependence of the percentage of $sp^2$ bonded atoms in
the region of the thermal spike on the heating ``temperature''
for different volumes.
When the thermal spike occurs in an initially amorphous sample the
structure of the damaged region after cooling exhibits the above 
dependencies as well as being a function of 
the structure of the
original amorphous carbon layers.

\end{abstract}


\maketitle

\section{\label{introduction:level1}  Introduction}

\begin{figure}
\includegraphics{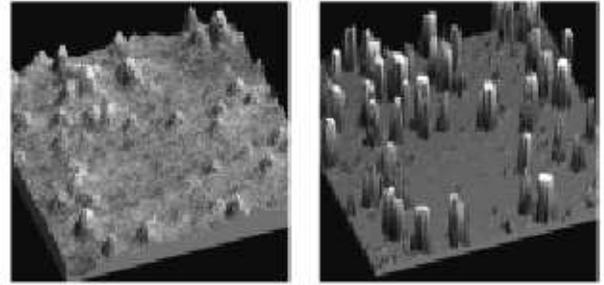}
\caption {\label{rafi3}
Surface morphology (left) and local electrical conductivity (right) 
of the transformed 
diamond-like amorphous carbon due to ion impact (taken from ref. 
\cite{hoffsass}).
}
\end{figure}

Carbon can bond in different hybridization forms, 
including diamond-like ($sp^3$)
 and graphite-like ($sp^2$), resulting in C based materials 
with  extremely 
different physical and chemical properties.  
After equilibrium conditions are achieved, 
disrupted $sp^3$ bonds may reconstruct as the more stable $sp^2$ bonds. 
The passage of swift ions through matter causes severe bond breakage, 
which is generally referred to as a ``thermal-spike''
that accompanies the stopping of the ion in matter. For the case of diamond,
the local bond disruption of the $sp^3$ bonds, and subsequent 
reconstruction upon equilibration, may result in the formation of an 
$sp^2$ rich 
region along the ion-track.  
Indeed, the passage of MeV or GeV heavy ions through
 diamond-like ($sp^3$ rich) layers has been shown by high resolution AFM 
and STM scans to result in local graphitization \cite{waiblinger}. 
This is evidenced by the 
formation of protrusions at the ion-impact points (swelling) 
and by the formation of electrically conductive channels. 
Both these factors indicate that the preferred $sp^2$ bonding 
has occurred along 
the pathways of the ions, in the otherwise $sp^3$ rich material 
\cite{hoffsass}.
 The width of the transformed material, due to the impact
 of GeV Au ions, has been deduced from the data an was found to be of 
the order of 85 nm. Fig. \ref{rafi3} illustrates 
the material 
transformation (density and electrical conductivity) 
measured by atomic force and scanning tunneling microscopy.

The damage inflicted on diamond by sub-MeV ion-implantation is known to 
result in electrical conductive regions along the ion track and around 
the stopping points of the ions \cite{dresselhaus}. 
The nature of these conductive regions, which have been shown to give 
rise to hopping conductivity, is controversial.  Prins postulates that they
  mainly contain a mixture of vacancies and interstitial C atoms which 
eventually collapse to from 
vacancy clusters ('vacloids') \cite{prins}.  
On the other hand, Prawer and Kalish have
concluded, based on measurements of the dependence of the
conductivity of damaged diamond on ion-damage-implantation dose
that the ion-affected diamond is highly conductive due to the
formation of graphite-like regions along the ion-tracks \cite{prawerkalish}.

 Computer simulations of the damage in diamond caused by the energetic 
displacement of a few C atoms resulting in damage to a small volume in 
diamond \cite{saada} (of 1.4
 nm$^3$) show that the size and nature of these regions depend on 
the initial recoil energy imparted to the C atoms and on the sample 
temperature. In some cases the formation of clusters of $sp^2$ bonded 
C atoms forming a six-fold rings has been observed in these simulations.

In the present study we simulate the passage of a swift ion through a diamond 
or a diamond-like tetrahedrally bonded amorphous C layer ($ta-C$) by  
locally ``heating''
 a section of the sample up to very high temperatures 
(comparable to the temperatures expected to prevail in the thermal spike).
We compute the $sp^2$ and $sp^3$ fractions in the affected
volume and follow their evolution as a function of cooling 
rate and time.
 We find that, upon cooling, a region 
which contains
 both $sp^2$ and $sp^3$ bonded C forms along the track. 
The nature of the material in this regrown region 
depends on the initial heating of the track and on 
its cooling rate. Under some conditions, complete 
regrowth back to perfect diamond is found, whereas in other cases, 
a $sp^2$ rich region remains along the ion track. 
Similar results are found when a thermal spike is simulated in 
amorphous carbon which is then allowed to cool down under 
a range of different conditions.  
The relevance of these computational results, obviously limited 
to small samples and to relatively short cooling times, to the actual ion 
passage through C based materials is discussed. 
 
We start below by briefly reviewing the physical processes
involved in the formation of the thermal spike and the typical
parameters of such a spike initiated in diamond by the passage of
a swift heavy ion; we then describe the computations performed,
discuss their relevance to ion-implantation experiments; present
and discuss our results.

\section{\label{background:level1} Background }
The ``thermal spike'' which accompanies the passage of an ion through 
matter can be viewed as a short-time local melting
of the ion damaged region followed by a rapid cooling. Quenching
from the liquid phase has been widely used to create amorphous
materials (including amorphous carbon) in computer simulations,
using empirical potentials \cite{tersoff}, {\it ab initio}
\cite{marks1,galli,alvarez} and tight-binding molecular dynamics
\cite{wangho, wangho1,frauenheim}. It has been shown that such a
procedure produces an amorphous carbon 
similar to material
produced experimentally by the deposition of a carbon film either
by evaporation in an electron beam, by carbon arc
\cite{li,kugler,kelires} or by ion beam deposition with ion
energies between several ten and several hundred eV
\cite{mckenzie,fallon,lifshitz}. 

In contrast to this large body of
computer simulations of amorphous carbon, there have been only a
few simulations of the material resulting from damaging diamond by
ion-implantation. Specific simulations of damage in material
induced by ion implantation can be divided into two categories:
(i) the commonly used TRIM code \cite{trim1,tomi} and (ii)
the Molecular Dynamics (MD) technique \cite{rapaport}. 
The TRIM (TRansport of Ions in
Matter) code is a Monte Carlo program that calculates the
trajectories of the primary ion and of the recoiling host atoms
involved in the irradiation event. The required input parameters are 
the ion type and energy and the properties of the host material
i.e. its composition, density, and the displacement energy of the
constituent atoms. TRIM provides good estimates of the damage and
of the implant distributions and it is widely used in planning the
dopant profile in ion-implantation experiments. However, TRIM
cannot account for the structure of the material (amorphous
vs. crystalline) nor model the dynamic annealing of the implantation
affected material. In particular, it cannot predict the nature of
the defects in diamond because of the variety of  possible bonding
configurations of carbon that may induce the diamond to graphite
transition. A modification of TRIM has recently been developed by
Sha'anan \cite{tomi} which enables following the evolution, in space and
time, of the damage cascade caused by the passage of a single ion
through matter. The results of the application of this code to
simulate the passage of an 1MeV Xe ion through diamond 
show that extremely high
temperatures (of the order of many tens of thousand of degrees)
are induced during very short times (less than 1 ps) along the ion track.
As we will discuss below, these resemble the time/temperature 
regime modeled in our computations of the damage inflicted on 
diamond and $ta-C$ by heavy ion irradiation.

An MD study of radiation damage caused by the
passage of a single carbon atom through diamond was performed by
Wu and Fahy \cite{wu}, using the Tersoff potential. In that
study the computations were restricted to the investigation of the
damage threshold energy necessary to displace a single C atom from
its lattice site. Smith \cite{smith} simulated the results of
the bombardment of diamond using a similar computational method
but limiting the analysis to the ejection mechanisms of C atoms
from a diamond surface (carbon self-sputtering). Marks et. al.
\cite{marks} have investigated the growth of $a-C$ layers. By computing
the mean square displacement of the atoms during the cooling of
the melt and the time required for it to relax, the 
life-time of the thermal spike in diamond was found to be less than 1
picosecond, for ion impacts below 400 eV. In these latter 
study {\it ab initio} MD techniques were used.

Saada {\it et al} \cite{saada} carried out MD
simulations, using the Tersoff potential, of a damage
region embedded inside a diamond matrix created by the energetic
displacement of carbon atoms with a kinetic energy of up to 416 eV
(8 times the displacement energy of carbon atoms in diamond) in
the bulk of a diamond crystal. Such regions are likely to appear in
diamond at the final stages of the implantation-related damage
cascade. The conclusion of that study was that partial
amorphization of a volume with a radius of 1-2 nm occurs around
the stopping point of the recoiling ion. With an increasing number
of successively displaced atoms, the damage volume and the number
of threefold coordinated atoms was found to increase. The
structure of the damaged regions (i.e. the ratio of $sp^2$ and $sp^3$
bonded C atoms) was found in ref. \cite{saada} to depend on 
the sample temperature.

These computer studies strengthen our understanding
of the processes occurring inside or at the end of an 'ion track'
in diamond. However, the transitions from $sp^3$ to $sp^2$ bonds
within thermal spike regions have not yet been investigated in detail.
The purpose of the present study is to investigate transformations of
diamond and diamond-like amorphous carbon under the extreme
conditions which prevail during and after the passage of a heavy
fast ion through the materials.

\section{\label{details:level1}  Tight-binding simulation details  }
 
\subsection{\label{comdet:level1}  General description of the 
calculations  }
 
We present a simulations of the thermal spike which occurs during
the passage of a heavy fast ion through either diamond or
amorphous carbon by locally heating a slab of the material to very
high temperatures and allowing it to rapidly cool down. 
A large
increase in volume accompanies the $sp^3$- to $sp^2$ bond
conversion \cite{prins1}, but the volume of the damaged region
in the center of the diamond is restricted by the surrounding diamond (or
amorphous carbon) lattice, therefore simulations are carried out
under constant volume conditions.
In order to set the temperature, a randomly directed velocity is assigned 
to each atom. The value of the velocity is randomly selected 
in accordance with a Maxwell 
distribution for the desired temperature. The velocities can be 
rescaled to be related to the ambient temperatures.     
Periodic boundary conditions are applied to 
the samples in all three directions. 
In order to describe the interactions between carbon atoms we use a tight-
binding model \cite{tightbinding}, which has been shown to be transferable to
 successfully simulate different carbon polytypes: diamond, graphite, 
fullerenes \cite{fuller} 
as well as disordered carbon structures such as liquid and amorphous 
carbon phases \cite{wangho,wangho1}. 
The MD step was 0.5 $\times$ 10$^{-15}$ s.

 The method used here to distinguish between $sp^3$ and $sp^2$ 
bonded atoms is
based on determination of the coordination number of each atom.
The radial distribution function (RDF) of all carbon samples
exhibits a clear gap, centered at about 0.19 nm, separating the
first and the second peak corresponding to the first and second
neighbor shells. All atoms within the sphere of radius 0.19 nm  
are thus assumed to comprise the nearest neighborhood of a
given atom. Therefore, the number of neighbors of each atom within
a distance of 0.19 nm determines the coordination number. 
 Atoms that have four nearest neighbors within this sphere are assumed to be
 $sp^3$ bonded, while atoms which have only three are assumed to be 
$sp^2$ bonded.

The Atomic Visualization package ({\bf AViz}) \cite{aviz} was used 
extensively in all stages of this project. 
A visualization of our carbon samples with color 
coding for different atomic bonding helped indentify 
clusters of either $sp^2$ or $sp^3$ coordinated atoms, 
rings, graphite-like planes. Color pictures of the results 
of the present study
can be found on the web site \cite{avizweb}.

\subsection{\label{gendet:level1}  Details of the sample preparation }

\subsubsection{\label{comdiam:level1} Damaged diamond
sandwiched between two layers of diamond}

Six different samples with a density of 3.5 g/cc, initially arranged 
as a perfect diamond crystal, were prepared. Their sizes were 
2 $\times$ 2$\times$ 6 (192 atoms), 2 $\times$ 2$\times$ 7 (224 atoms),
2 $\times$ 2$\times$ 8 (256 atoms), 2 $\times$ 3$\times$ 6 (288 atoms) 
and 3 $\times$ 3$\times$ 6 (432 atoms) diamond unit cells.

\begin{figure}
\includegraphics{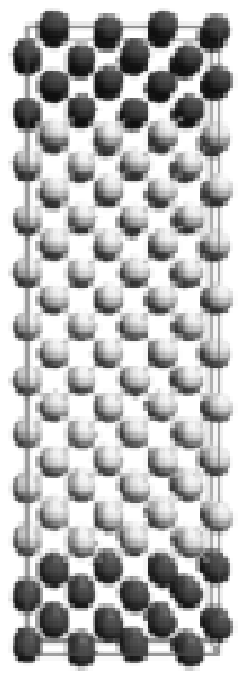}
\caption {\label{init}
Initial configuration of the samples; grey (purple) balls 
represent the frozen atoms,
white (yellow) balls represent atoms that are free to move.
(Purple and yellow are shown online)}
\end{figure}

The ``thermal spike'' was created by heating the central layers 
up to ``temperatures'' 
of 1.2-2.6 eV (14000-30000 K). 
The 32 upper and 32 lower atoms of each sample (the upper and lower layers 
of a height of 3.55 \AA) were frozen, i.e. 
the motion of these atoms was forbidden.
The initial geometry of the samples is shown in Fig. \ref{init}.
The outermost frozen diamond layers had the effect on their 
neighboring hot atoms of
forcing them to return to their initial positions i.e. to regrow
epitaxially back to diamond.
In this way a temperature 
gradient from the edges to the center of the sample was created.
Once the liquid phase reached equilibrium (this process was controlled 
by monitoring the total energy of the system),
the central layers were cooled 
to room temperature at a cooling rate of 10 K/fs.
The height of the hot layers  
varied between 4 and 6 diamond unit cells and 
the width of the hot layers  
varied from 2 to 3 diamond unit cells.
The full time of cooling was 1.5 ps (3000 MD steps). 
However, the central layers ceased to move after the 750 first MD steps, 
so the lifetime of our ``thermal spike'' was 0.3 ps. This value is similar to 
the estimates of Marks \cite{marks} and Saada \cite{saada}, 0.2ps and 0.21ps,
respectively. 

\subsubsection{\label{comamor:level1} Damaged amorphous C region 
sandwiched between two layers of frozen $ta-C$}

Four initial samples of diamond-like amorphous carbon 
at density 3.5 g/cc were 
prepared by heating diamond up to 5000 K followed by cooling 
of the liquid phase to 300 K during 5 ps with an exponential 
cooling profile. 
Each $ta-C$ amorphous sample contained 192 atoms with a geometry 
7.1 \AA $\times$ 7.1 \AA $\times$ 21.3 \AA.
The samples contained  
66$\pm$4 \% of $sp^3$ coordinated atoms.
In order to achieve a larger percentage of $sp^3$ coordinated atoms 
(above 70 \%), two additional samples were prepared using the 
the following procedure: the amorphous structure was prepared 
with a density of 4.0 g/cc by quenching a liquid phase down from 
10000 K to 700 K with a cooling rate of about 500 K/ps, then the 
density was reduced from 4 to 3.5 g/cc by uniform expansion,
after which the temperature of the system was raised to 2500 K,
and thereafter the structure was recooled to room temperature 
\cite{wangho1}.
The two structures obtained at a final density of 3.5 g/cc 
consisted of 84 \% and 77.5 \% of $sp^3$ bonded atoms 
respectively.  
 
To simulate a ``thermal spike'' in the diamond-like 
amorphous carbon samples, the central region of each sample 
was heated to ``temperatures'' of 1.2-2.60 eV per atom (14000-30000 K). 
The upper and lower layers of a height of 3.55 \AA $ $ each,
were kept frozen.
For the six amorphous carbon samples the height of the hot layers  
was 14.2 \AA $ $ (128 atoms in average).
A representation of a typical initial sample geometry is given 
by Figure \ref{init}. Dark atom at top and bottom
indicate the frozen layers, the light central ones 
are free to move and when heated, there is  
a temperature gradient from the edges to the center of the samples.
Following the heating stage 
the structures were cooled to room temperature at a cooling 
rate of 10 K/fs. The time of calculations and the lifetime 
of the ``thermal spike'' were the same those described above for 
the case of the damaged 
region between two layers of perfect diamond.

\subsection{\label{equilibration:level1}  Sample equilibration} 
  
We checked carefully, by monitoring of the 
total energy of the system, that
the heating time was sufficiently long to enable the 
liquid structure to reach equilibrium 
(early attempts to prepare samples of amorphous carbon 
by quenching from a liquid carbon that was not in 
equilibrium led to inconclusive results). 

\subsection{\label{relevance:level1}  Relevance to experiment} 

Our aim is to determine the nature of transitions between 
different carbon bonds  as the result of the passage of fast ions
 in samples of different initial structures. To do this, we need to
examine the region of passage and its intermediate neighbourhood
 on an atomistic scale. However, this microscopic region 
is contained in a larger carbon matrix, and while the microscopic
region corresponds directly to that in the experiment, 
the surrounding matrix may differ in scale.
We address the issues of system size and temperature
in relation to the parameters of experimental samples.

In the simulations,
on one hand,  a maximum of 
432 atoms can be contained in a single sample
due to computational limitations. 
On the other hand,  the periodic boundary conditions mean that
the sample is in a sense infinite, and, in particular, has no surfaces.
Thus the simulation samples are both smaller and larger than the
 laboratory experiment.
We are investigating  the  local region around the ion path
(which has a linear extent of 7.2 \AA $ $ along the track directon
 prior to repetition). 
The geometry was chosen so that this region is well surrounded,
 in the direction perpendicular to the track, by 
diamond or amorphous carbon matrix, which is frozen in place.
 In particular, there are 7.4 \AA $ $ between replicas of the tracks, 
a region that is sufficient to avoid interaction since the range of
 the potential is only 3 \AA. In order to ensure that the samples are of 
sufficient size to enable reliable interpretations, a series of different
 sizes were used for each situation and the effect of the scaling of the
 system size was carefully investigated. 
Although this is not exactly equivalent to the finite size scaling 
\cite{binder}
that is commonly applied to phase transitions and critical phenomena,
and is closer to the scaling used by Rosenblum et al \cite{irina}
in a study of thermal stress at diamond interfaces. The 
excellent agreement of both of these scaling approaches with results
from laboratory experiments
gives us confidence in the concept.
The collapse, as a function of system size, and the
smoothness  of the  scaling that will be demonstrated below confirms 
that the results can be reliably extrapolated even further if needed.

Temperature is introduced into the simulations as kinetic energy.
In the laboratory, the region around the ion track is extremely hot 
for a very short time. The high heating ``temperatures'' of the simulation 
are thus large amounts of kinetic energy imported the heated atoms. 
Here, too, we have studied 
the scaling of ``temperature'' with system size.
Not surprisingly, the actual ``temperature'' needed to attain specific
 structures is a function of system size. Again, collapse of graphs 
to a single functional form  will be demonstrated and provides  
validation of our approach.

\section{\label{results:level1}  Results of tight-binding simulations }

\subsection{\label{diamond:level1} Damaged diamond sandwiched  between 
two layers of intact diamond }

\subsubsection{\label{temperature:level1} Influence of the heating 
``temperature''} 

When the diamond lattice in the central hot layers was
melted, $sp^2$ and $sp$ bonded atoms appeared 
in the liquid layers. Upon cooling some (or all) of 
these $sp^2$ and $sp$ bonded atoms 
transformed back to $sp^3$ coordinated C atoms. 
The final fraction of threefold and twofold coordinated 
atoms depends on the kinetic energy per atom (``temperature'') 
in the heating process. The higher the heating ``temperature'' of
the atoms the larger was the
ratio of $sp^2/sp^3$, that prevailed after cooling. 
Samples initially heated to a 
relatively low ``temperature'' returned to diamond, whereas 
for hotter samples some non diamond bonds remained after cooling.

\begin{figure}
\includegraphics{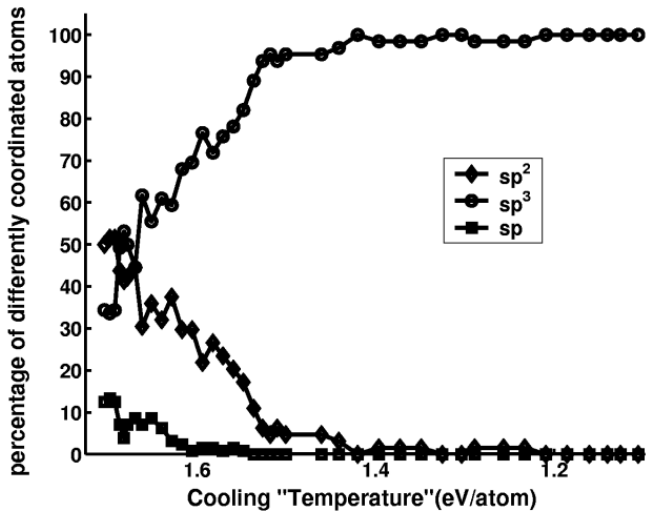}
\caption{ \label{perc141}
Recovery rate showing the evolution of the fraction of two-, 
three- and fourfold coordinated atoms in 4 layers (128 atoms),
following heating to 1.725 eV and subsequent cooling. } 
\end{figure}

For example, let us consider the 2 $\times$ 2$\times$ 6 sample 
containing 192 atoms, 128 of which were free to move 
(i.e. were heated and subsequently cooled). 
The percentages of four-, three-, and twofold 
coordinated atoms in the hot layers after cooling from different 
``temperatures'' are presented 
in Table \ref{tab141}. Whenever the heating ``temperature'' was lower 
than 1.72 eV per atom 
all $sp^2$ and $sp$ bonds disappeared in the process of cooling,
and the final, cooled, structure returned to a perfect diamond lattice,
as shown in  Fig. \ref{perc141} where the percentages of different 
hybridizations are graphed as a function of cooling ``temperature''.

\begin{figure}
\includegraphics{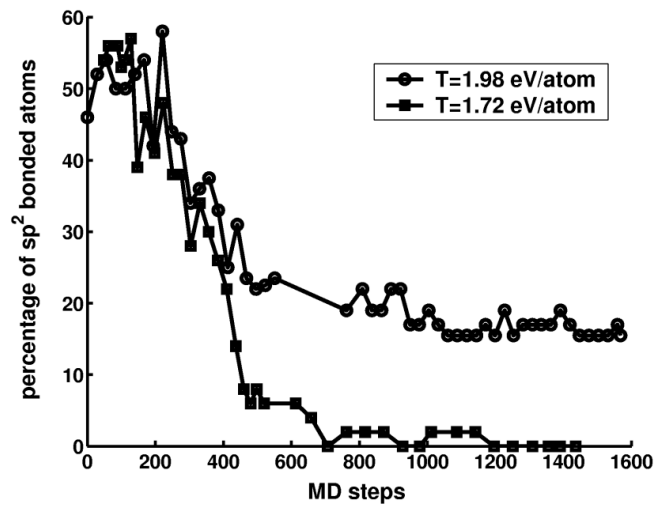}
\caption{ \label{perc141_1}
 The fraction of threefold coordinated atoms in the hot layers, shown during
the process of cooling from the initial ``temperature'' of 1.725 eV 
and 1.98 eV/atom. 
The sample contained 4 hot layers (128 atoms). } 
\end{figure}

At higher initial heating ``temperatures'' the heated 
layers remained, after cooling, partially or 
entirely amorphous. The fraction of three- and two-fold coordinated 
atoms in the structure are shown in Fig. \ref{perc141_1} which shows the 
percentages of $sp^2$ bonded atoms during the cooling process
from different initial heating ``temperatures''.

\begin{table}
\begin{center}
{\small
\begin{tabular}{|c|c|c|c|} \hline

$T$ (eV/atom) & Fourfold(\%) & Threefold (\%) & Twofold (\%)\\ \hline 
1.55 eV & 100 & 0 & 0 \\ \hline
1.72 eV & 95$\pm $5 & 5$\pm $5 & 0 \\ \hline
1.98 eV & 79$\pm $5 & 19$\pm $5 & 2$\pm $2\\ \hline
2.41 eV & 74$\pm $5 & 23$\pm $5 & 3$\pm $2\\ \hline
2.58 eV  & 62$\pm $5 & 34$\pm $5 & 4$\pm $3\\ \hline

\end{tabular}
}
\end{center}
\caption{Fraction of the four-, three-, and twofold 
coordinated atoms in the heated layers after cooling in 
the sample with 192 atoms.}
\label{tab141}  
\end{table}

\begin{figure}
\includegraphics{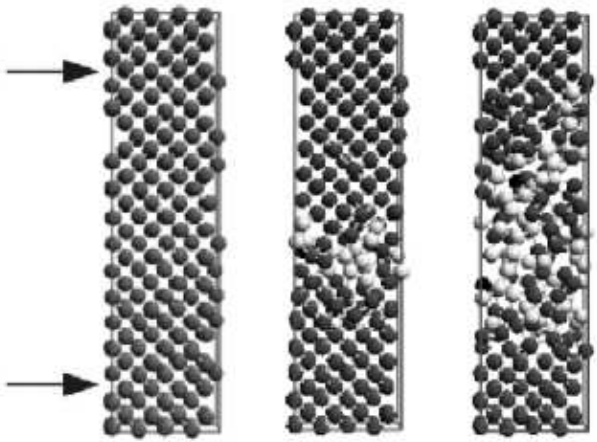}
\caption {\label{tmuna161}
 Amorphous carbon located between two layers of diamond.
The samples contained 192 atoms (2 $\times$ 2$\times$ 6), 
temperature of heating was 1.2 eV/atom (left), 1.38 eV/atom (center) and 
1.98 eV/atom (right). 
Grey (purple) balls represent fourfold 
coordinated atoms, white (yellow) balls represent 
threefold coordinated atoms, black balls (which appear 
only in the right most sample) represent 
twofold coordinated atoms. Boundaries between frozen layers 
and layers of atoms which are free to move are denoted by arrows.
(Purple and yellow are shown online)
}
\end{figure}

The volume of the region remaining amorphous after cooling 
was found to increase 
with the heating ``temperature''. In Fig. \ref{tmuna161}  
visualization of some samples cooled from different heating 
``temperatures''are presented.
We see that in the sample where the heating ``temperature''
was high (1.98 eV/atom) almost an entire layer of atoms which 
were free to move has transformed to amorphous carbon 
with 19 \% of $sp^2$ 
bonds, while in the sample initially heated to 1.38 eV/atom
only a small central region is amorphous and 
the layers which are adjacent to the frozen part of the sample 
return to their crystalline diamond structure.
The reason for the dependence of the volume that
remains amorphous on heating ``temperature'' is that 
at lower ``temperatures'' the frozen outer layers of diamond extend 
their influence to 
larger distances, therefore a larger volume of the damaged region 
reconstructs to the diamond structure after cooling.   
A tendency to segregation of the threefold and the fourfold atoms was  
observed in the samples heated to high temperatures when 
the volume of the region that remained amorphous was 
sufficiently large (almost all non-frozen atoms).

\begin{figure}
\includegraphics{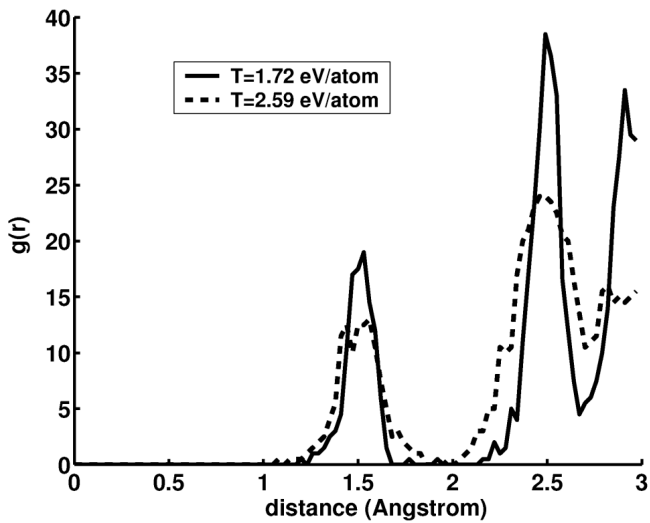}
\caption {\label{rdf141}
Radial distribution function g(r) of central layers of the 
sample containing 4 hot layers (2 $\times$ 2$\times$ 6) 
cooled from two different initial ``temperatures''.
}
\end{figure}

The radial distribution function (RDF) of the C atoms in the hot 
layers is presented in 
Fig. \ref{rdf141}. The second peak is visible in the 
figure due to the layers which have reconstructed to 
diamond structure after cooling. However, as was explained above, 
the higher the ``heating temperature'' the larger the volume of the
region remaining amorphous. Therefore all RDF peaks 
related to the higher heating ``temperature'' (dashed line) are lower 
and broader.
For high heating ``temperatures'' the overlapping peaks are 
a superposition of two peaks that
correspond to $sp^2$ bonded and $sp^3$ bonded atoms.
  
\begin{figure}
\includegraphics{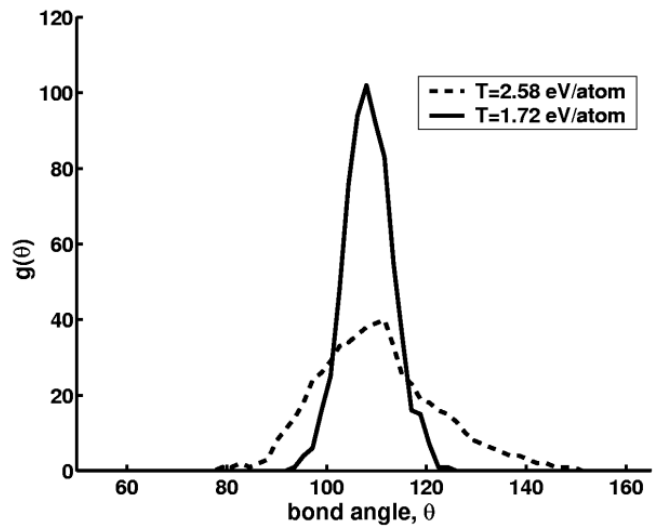}
\caption {\label{adf141}
Angular distribution function of central layers of the sample 
containing 4 hot layers (2 $\times$ 2$\times$ 6) cooled 
from two different initial ``temperatures''.
}
\end{figure}

The angular distribution function of the hot layers demonstrates 
similar behavior, see Fig. \ref{adf141}.
The peaks are lower and broader when the heating ``temperature'' is 
higher, this corresponds to a higher fraction of three- and twofold 
coordinated atoms.
The angular distribution peak related to $T=$1.98 eV is located at 112$^\circ$ 
(closer to the ideal 120$^\circ$ for $sp^2$ bonded C atoms),
while the peak of the curve related to 1.72 eV is located at 109$^\circ$
(typical for $sp^3$ bonded atoms).

\subsubsection{\label{size:level1}Effects of the size of the hot layers.}
The heating ``temperature'' is not the only factor that influenced 
the structure of the damaged region.
If the height of the layer with atoms that were free to move 
was increased from 4 to 6 unit cells without changing the width of samples, 
the percentage of three- and twofold coordinated 
atoms in the damaged layers, when heated to the same temperature 
also increased (see Table \ref{diffsize}).
As was explained above, the frozen layers assist the 
intermediate layers in their vicinity to return to diamond lattice 
structure. 
The central atoms in the samples with a large 
number of the hot layers feel the influence of the diamond 
frozen layers to a lower extent, 
and therefore the volume of the region remaining amorphous after cooling 
was large, with a high percentage of $sp^2$ coordinated atoms. In the small 
samples the central layers needed to be heated to very 
high ``temperatures'' to diminish the influence of the 
outer frozen layers. The highest ``temperature'' at which the 
sample reconstructed 
its diamond lattice after cooling also depended on the 
height of the hot layers. This temperature decreased as 
the height of the hot layers increased (see Table \ref{tempdiam}).

\begin{table}
\begin{center}
{\small
\begin{tabular}{|c|c|c|c|} \hline

Number of layers & Fourfold(\%) & Threefold (\%) & Twofold (\%)\\ \hline 
4 & 95$\pm $5 & 5$\pm $5 & 0 \\ \hline
5 & 78$\pm $5 & 21$\pm $5 & 1 \\ \hline
6 & 74$\pm $5 & 26$\pm $5 & 0-2\\ \hline
\end{tabular}
}
\end{center}
\caption{Fraction of the four-, three-, and twofold 
coordinated atoms in the samples 
with different height of the hot layer heated to 1.72 eV per atom.}
\label{diffsize}  
\end{table}     

\begin{table}
\begin{center}
\begin{tabular}{|c|c|} \hline

Number of layers& $T^*$\\ \hline 
4 & 1.63 $\pm $0.04 eV \\ \hline
5 & 1.51 $\pm $0.04 eV \\ \hline
6 & 1.38 $\pm $0.04 eV \\ \hline
\end{tabular}
\end{center}
\caption{The upper ``temperature'' of heating at which the samples 
reconstruct their diamond structure ($T^*$).}
\label{tempdiam}  
\end{table}     

\begin{figure}
\includegraphics{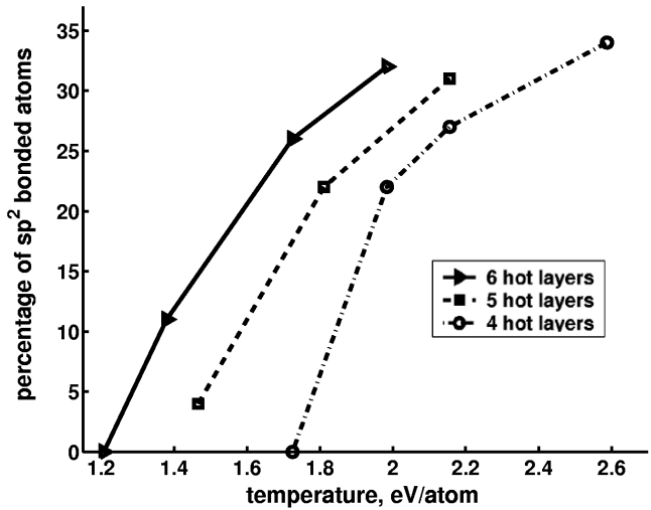}
\caption {\label{percent1} The dependence of the percentage 
of $sp^2$ bonded atoms 
on the heating temperature for different heights of the hot layers.}
\end{figure}

\begin{figure}
\includegraphics{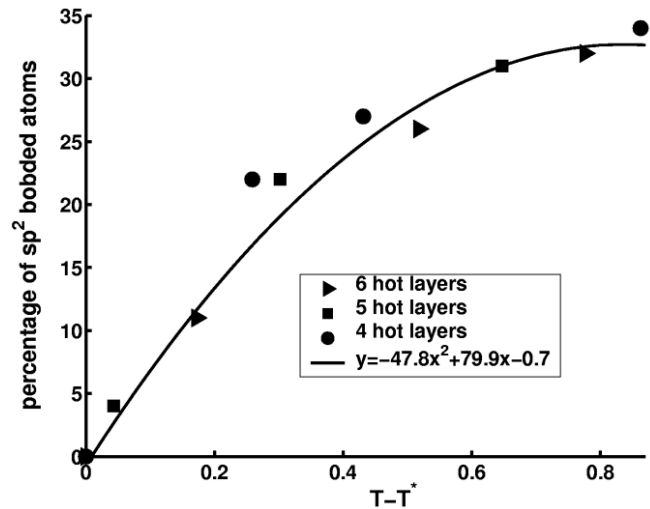}
\caption {\label{percent2} The same curves shifted to $T^*$. The solid line is 
a result of square polynomial fitting.}
\end{figure}

Graphs of the dependence of the percentage of $sp^2$ bonded atoms 
on the heating ``temperature'' are plotted in Fig. \ref{percent1} 
for the samples with different heights and same width of the hot layers.
A striking feature of these curves is that they have 
the same shape for the different heights 
of the hot layer. All the curves can be fitted to a unique 
function $f(T-T^{*})$, where $T^*$ is the upper temperature of heating, 
at which the sample can return to diamond after cooling. 
In Fig. \ref{percent2} the collapse of the curves 
to a single scaling function is shown.

The samples with different width (from 32 to 72 atoms in each layer) 
at the same height of 
the hot layers also demonstrate finite size effects.
The percentage of three- and twofold coordinated atoms in the central layers 
is larger in the wider samples at the same heating ``temperature'', 
the ``temperature'' at which the sample returns to diamond, $T^*$, is higher 
for the thinner samples (see Table \ref{thick}).

\begin{table}
\begin{center}
\begin{tabular}{|c|c|c|c|} \hline

Width & $T^*$ & 1.72 eV & 2.07 eV\\ \hline 
32 atoms & 1.63 $\pm $0.04 eV & 5 $\pm $5 & 20 $\pm $5 \\ \hline
48 atoms & 1.48 $\pm $0.04 eV & 19 $\pm $5 & 32 $\pm $5 \\ \hline
72 atoms & 1.29 $\pm $0.04 eV & 22 $\pm $5 & 41 $\pm $5\\ \hline
\end{tabular}
\end{center}
\caption{The upper ``temperature'' of heating at which the sample 
reconstructed its diamond structure ($T^*$) and percentage of $sp^2$ 
coordinated atoms in the samples cooled from 1.72 and 2.07 eV per atom.}
\label{thick}  
\end{table}

\subsection{\label{amorphous:level1}Simulation of molten and 
cooled amorphous carbon 
located between two frozen layers of amorphous carbon.}

\begin{figure}
\includegraphics{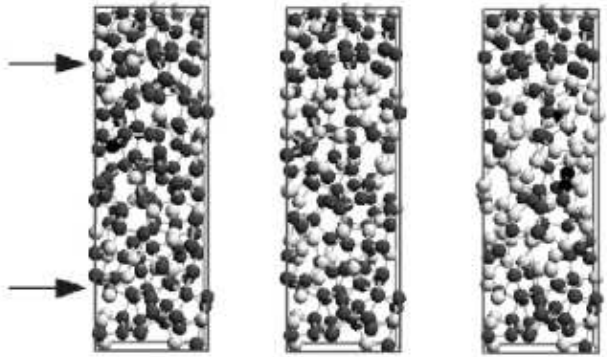}
\caption {\label{amam} Samples of amorphous carbon located between 
two layers of $ta-C$ amorphous carbon: the initial sample 
(77.5 \% of $sp^3$) (left), and the samples after cooling 
from 2.07 eV (in the center), and from 2.33 eV (right). 
Grey (purple) balls represent fourfold 
coordinated atoms, white (yellow) balls represent 
threefold coordinated atoms, black balls (of which there are only four 
in the right most sample) represent 
twofold coordinated atoms. Boundaries between frozen layers 
and layers of atoms which are free to move are denoted by arrows. }
\end{figure}

Now let us consider structure of a central amorphous carbon layer 
sandwiched between two 
layers of frozen amorphous carbon. After cooling,   
the central damaged 
region remained amorphous, but the $sp^2/sp^3$ ratio changed.  
The dependence of this generated structure on the heating ``temperature''  
was similar to that of the previous stage of the simulation (amorphous 
carbon between two layers of diamond). Namely, if the central atoms 
were heated 
to higher ``temperatures'', the fraction of the $sp^2$ bonded atoms 
increased (correspondingly, the fraction of the $sp^3$ bonded atoms 
decreased). Some of these structures 
were highly inhomogeneous, in particular when the initial sample 
contained a large fraction of fourfold bonded atoms and was heated 
to a relatively high ``temperature''. In Fig. \ref{amam} the 
visualization of some of these amorphous carbon samples is presented. 
We can see that 
the frozen layers of the sample with 77.5 \% of $sp^3$ bonded atoms 
initially, contained after heating and subsequent cooling 
75 \% of $sp^3$ bonded atoms, while the central 
layers after heating to 2.33 eV and subsequent quenching 
contained only 39 \% of the fourfold atoms.

\begin{figure}
\includegraphics{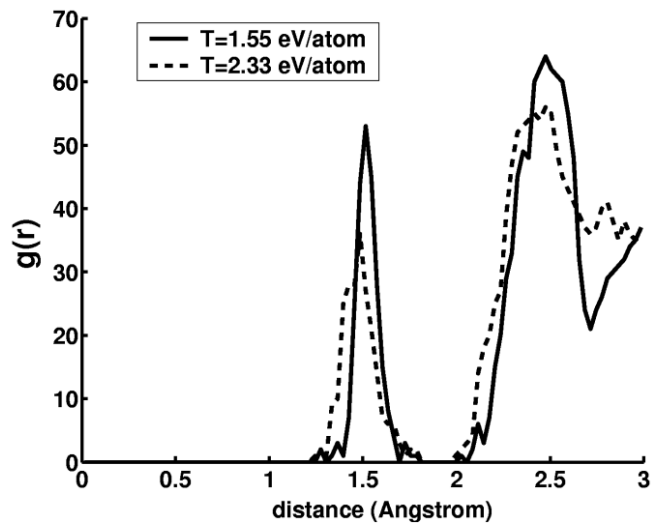}
\caption {\label{rdfamor}
Radial distribution function $g(r)$ of central layers of the sample 
containing 77.5 \% of $sp^3$ coordinated atoms initially  
(79 \% of $sp^3$ in the central layers) cooled 
from two different ``temperatures'': after cooling from 1.55 eV the 
central layers contained 78 \% of $sp^3$ coordinated atoms, 
while after cooling from 2.33 eV the 
central layers contained 39 \% of $sp^3$ coordinated atoms.
}
\end{figure}

\begin{figure}
\includegraphics{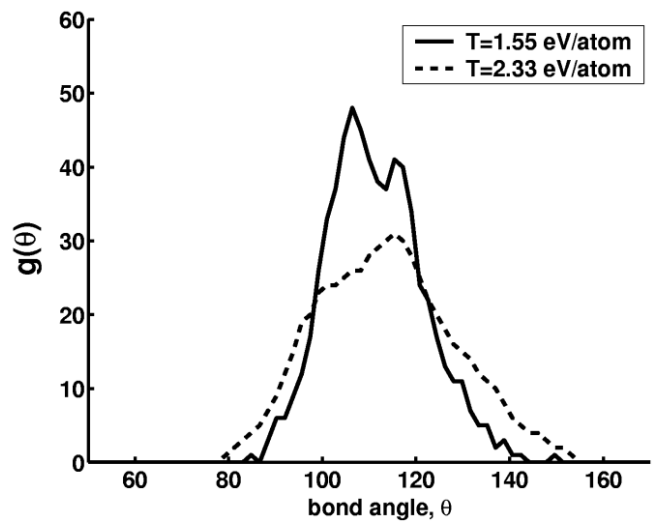}
\caption {\label{adfamor}
Angular distribution function of central layers of the sample 
containing 77.5 \% of $sp^3$ coordinated atoms initially  
(79 \% of $sp^3$ in the central layers) cooled 
from two different ``temperatures''.
}
\end{figure}

The radial (RDF) and the angle (ADF) distribution functions of 
the central layers of this amorphous carbon 
sample after cooling from the ``temperatures'' of 1.55 eV and 2.33 eV 
are plotted in Figs. \ref{rdfamor} and \ref{adfamor} respectively.
The peaks corresponding to the higher temperatures of heating were 
lower, broader, and slightly shifted toward the shorter, 
graphitelike, bond length and larger, graphitelike, bond angle relative 
to those at the lower temperatures.
This indicates a higher fraction of the $sp^2$ bonded atoms.

\begin{figure}
\includegraphics{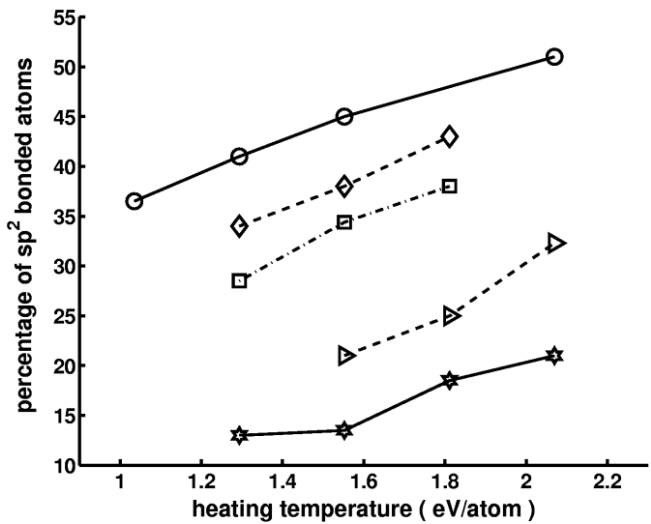}
\caption {\label{graphamor} The final percentage of $sp^2$ coordinated 
atoms as a function of heating ``temperature'' plotted for samples 
with different percentage of $sp^3$ and $sp^2$ bonded atoms before heating 
of the central layers. Circles in solid line-38 \% of $sp^2$ in initial 
amorphous sample (36\% in the central layers), diamonds in 
dashed line-36\% of $sp^2$ in initial 
amorphous sample
(31\% in the central layers), squares in dash-dot line - 30\% of 
$sp^2$ in initial 
amorphous sample (27.5\% in the central layers), triangles in dashed line - 22.5\% 
of $sp^2$ in initial amorphous sample (21\% in the central layers)
stars in solid line - 16\% of $sp^2$ in initial 
amorphous sample (14\% in the central layers). 
}
\end{figure}

It is obvious that the percentage of differently coordinated atoms in 
the frozen layers has influence on the results of the central region. 
The samples initially containing 
a larger fraction of the $sp^2$ bonded atoms, contained a larger fraction 
of the $sp^2$ bonded atoms after cooling (at the same temperature of heating).
Fig. \ref{graphamor} shows the final percentage of $sp^2$ coordinated 
atoms as a function of heating ``temperature'' plotted for samples 
with different percentage of $sp^3$ and $sp^2$ bonded atoms before heating 
of the central layers. The data related to the smaller initial fraction 
of $sp^2$ atoms have a lower final fraction of the $sp^2$.

\subsection{\label{gap:level1}The band gap}
\begin{figure}
\includegraphics{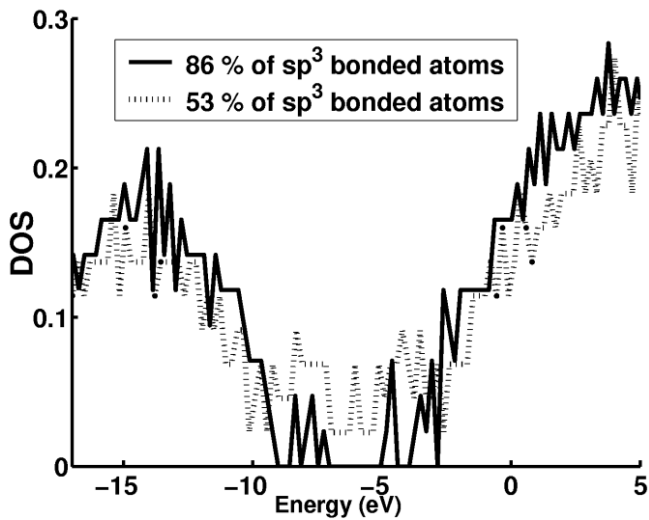}
\caption {\label{dos} The density of states of the samples of 
amorphous carbon, containing 86 \% of $sp^3$ bonded atoms 
and 53 \% of $sp^2$ bonded atoms. ($E=0$ is chosen arbitrary.)}
\end{figure}

\begin{figure}
\includegraphics{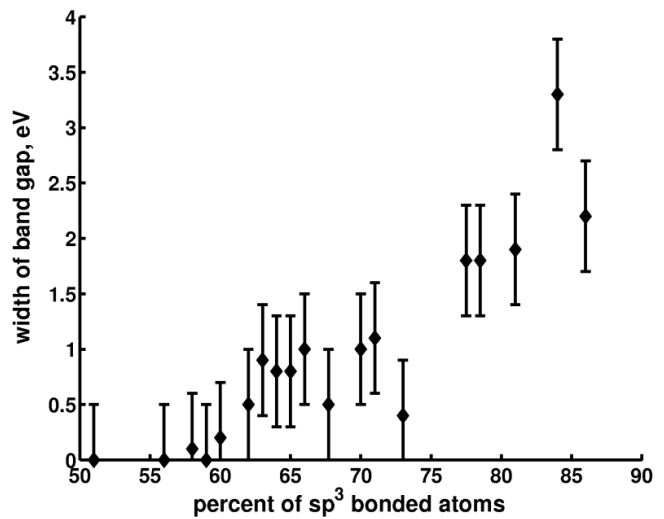}
\caption {\label{finalbandgap} The width of the band gap as a function of 
the percentage of $sp^3$ bonded atoms. The error bars represent 
an unaccuracy in visual determination of the band gap from DOS.}
\end{figure}

During the calculations many samples of amorphous carbon 
with different structures were generated. The width of the 
band gap was ``measured'' for most of them by computing of 
density of states (DOS) as function of energy. 
For example, the graph of density of states for two samples 
of amorphous carbon, containing 86 \% of $sp^3$ bonded atoms 
and 53 \% of $sp^3$ bonded atoms is plotted in Fig. \ref{dos}.
The bandgap of the sample with 86 \% of $sp^3$ bonded atoms 
was estimated visually from the graph to be 2.3 eV. 
The sample with 53 \% of $sp^3$ bonded atoms does not indicate 
the presence of any band gap, this sample is expected to be 
electrically conducting. The width of the band gap 
depends not only on 
the fraction of the differently coordinated atoms, but also on their 
individual 
configuration, i.e on the local clustering of $sp^2$ bonded atoms.
Fig. \ref{finalbandgap} shows a plot of the width of the bandgap as 
a function of the fraction of $sp^2$ coordinated atoms
 in various $ta-C$ samples.

There is a noticeable scattering of the points between 60 and 
75 \% of $sp^3$ bonded atoms. We suspect that in this interval the fraction 
of $sp^2$ bonded atoms increases sufficiently to begin to create clusters. 
If the fraction of the $sp^2$ atoms is smaller than 25 \%, 
than only separated $sp^2$ bonded atoms are embedded in the 
$sp^3$ amorphous 
network. If the fraction of the $sp^2$ atoms is larger than 40 \%, 
the $sp^2$ clusters begin to connect one with another. 
The band gap disappears 
when the fraction of the $sp^2$ bonded atoms reaches 45 \%.

\section{\label{sec:level1}Discussion and comparison with experiment.}

The experimental results  \cite{mcculloch,prawerkalish,waiblinger}
described above show that the the high local energy 
deposition of the ions into diamond or amorphous carbon 
along their paths causes a rearragement of the carbon atoms 
and can lead to a transformation of the insulating 
diamond or amorphous 
diamond-like $sp^3$ form of carbon into a conducting 
layer rich in $sp^2$ bonded carbon atoms.
In the simulations we modeled the transformation 
of highly ordered diamond structures into amorphous structures with 
the presence of graphitic like $sp^2$-bonds. In the samples which
were initially amorphous, rearragement of the amorphous structure and an  
increase of the fraction of $sp^2$ bonded atoms has occurred.

The experiments  \cite{mcculloch,prawerkalish} show that the 
resistivity, both in diamond and in amorphous carbon networks,
decreases with ion dose. In diamond there exists a critical 
damage density 
$D_c$ below which the diamond restores this pristine structure 
after annealing \cite{critical}. At $D=D_2$ a sharp drop in 
resistivity is observed and at higher doses graphitic conduction 
is measured in 
the irradiated volume. Kalish et $al$ \cite{percolation} investigated 
the possibility that the sharp drop in resistivity 
could be explained in terms of a percolation 
transition between conducting islands 
produced by the impinging ion, but found that 
the $R$ vs $D$ dependence was not sharp enough 
to be consistent with such a simple percolation theory. 
By contrast,
Hoffman et $al$ \cite{hofman} in their experiment 
on 1keV Ar irradiated diamond proved that the amorphization 
is a very sudden process and at $D_2$ the diamond collapses to 
an amorphous but still highly insulating form similar 
to amorphous diamond ($ta-C$). Only at doses exceeding those 
required for amorphization, does the amorphous $sp^3$ structure 
transform to an amorphous $sp^2$ structure which is electrically 
conducting. In our simulation at low ``heating '' temperatures 
that corresponds to a slower or lighter ion impact, the diamond 
also reconstructs this structure after cooling. 
At higher heating ``temperatures'' the structure remains amorphous 
after cooling and the fraction of $sp^2$ bonded atoms in the damaged 
region increases with the heating ``temperatures''. The faster the 
ion the larger the damaged region. In our simulation increasing 
of the heated volume leaded to increasing of  $sp^2$ fraction 
in this ion track and to increasing of conductivity.
   
Experimental results \cite{mcculloch} show that 
in contrast to diamond , for $ta-C$ where no crystalline template 
exists, there is nothing to inhibit the $sp^3$ to $sp^2$ 
bond conversion, hence there is no reason to expect a critical dose
at which a sharp increase in the electrical conductivity should start. 
Moreover, the decrease of the electrical resistivity 
in $ta-C$ occurs at much lower implantation dose than in diamond 
\cite{mcculloch}.
The results of electron energy-loss spectroscopy allows 
to estimate the fraction of $sp^2$ bonds, $f$, in ion implanted $ta-C$.
At low ion doses the $f$ increases linearly with 
ion dose, at a dose 10$^{14}$ ions/cm$^3$ ,$f$ becomes less sensitive 
and increases only slightly with dose.   
In our calculation $ta-C$ becomes conductor at much lower 
heating ``temperatures'' than diamond, this means that in 
contrast to $ta-C$, diamond needs 
an additional energy to break its crystal lattice and 
lead to amorphization upon cooling.

The conductivity in our amorphous carbon samples 
deduced from the shrinkage of the band gap 
increases with percentage of $sp^2$ bonded atoms in the damaged region. 
There are some slight discrepancies with experimental studies of 
electronic structure of damaged region. For example, 
McCulloch {\it et al} \cite{mcculloch} 
found a band gap 
of 2.5 eV at 85 \% of $sp^3$ bonded atoms, whereas 
we find that a value of 3.3 eV 
corresponds the fraction of 84 \% of the $sp^3$ atoms.

\section{\label{sec:level1}Summary}
We have simulated a damaged region
occuring in diamond and in diamond-like amorphous carbon by 
local heating of a slab in the sample of thickness comparable to that created 
by the ``thermal spike'' due to the passage of an energetic ion  
followed by quenching back to room temperature. 
The ``temperature'' of the heating varied from 1.2 to 
2.59 eV/atom (14000-30000 K).
For the case of diamond we found that at a heating ``temperature'' 
higher than a characteristic temperature
$T^*$ a small region in the center of the sample remains amorphous.
The size and the structure of the amorphous region depends on the 
heating ``temperature''; the size and the fraction of three- and 
two-coordinated atoms in the region increase with the heating 
``temperature''. When we increased the number of the free-to-move layers, 
the fraction of three- and 
two-coordinated atoms in the central layers increased. 
The curves of the dependence of the fraction of the $sp^2$ bonded atoms 
in the central layers on the temperature of heating have 
similar shape  
for different sizes of the hot layers.  

The same procedure was applied to simulate a damaged region in amorphous 
carbon samples. The central layers of each sample were melted at 
a very high temperatures (1.29-2.33 eV/atom) and then cooled by keeping 
the outer layers frozen. The percentage of three- and twofold 
coordinated atoms in the obtained 
damaged region was found to increase with heating ``temperature'' and 
with the initial fraction of the threefold coordinated atoms
in the sample.
Most of the structures of amorphous carbon generated in different 
stages of our simulation were highly inhomogeneous and clustering 
of $sp^2$ bonded C atoms has been observed. 

The band gap measured for all the amorphous carbon samples 
became narrower when the population of $sp^2$ bonded atoms 
increased. It vanished when the fraction of $sp^2$ bonded atoms 
reached a value of 45 \%. Beyond this fraction the damage 
affected volume becomes conductive.  

The results obtained in this study support the experimental 
conclusions that in the damaged region formed after 
the passage of a heavy fast ion either through diamond or amorphous carbon,
amorphization and sometimes conversion of $sp^3$ bonds to 
$sp^2$ bonds occurs. Another conclusion which follows 
from the simulations is that after passing of a faster (or heavier) ion 
(that corresponds to higher temperature of the ``thermal spike''),
the damaged region is larger and more $sp^3$ bonds are transformed to $sp^2$.
Although the present simulations do not exactly reproduce the conditions 
of the experimental ``thermal spike'', qualitatively 
our results 
are in good agreement with experiments.

\begin{acknowledgments}
We are  grateful to  Dr. G. Wagner  and Dr. D. Segev 
(formely known as Saada) for their contribution
to this project. We thank Prof. A. Horsfield and Prof. M. Finnis for providing us 
with the OXON package.
This work was supported in part by the Israel Science Foundation 
and the German Israel Foundation (GIF).
\end{acknowledgments}


\end{document}